\documentclass[10pt]{article}

\usepackage{lineno}
\usepackage{rotating}
\usepackage{hyperref}
\hypersetup{backref,colorlinks=true}

\def\Sq{\textit{Sq }} 
\def\Sqq{\textit{Sq}$^0$}

\begin{document}

\title{\textbf{Solar forcing of the terrestrial atmosphere}}

\author{Thierry Dudok de Wit$^a$, J\"urgen Watermann$^{a,b}$}

\date{\small $^a$ Laboratoire de Physique et Chimie de l'Environnement et de l'Espace, \\UMR 6115 CNRS - Universit\'e d'Orl\'eans, 3A avenue de la Rechecherche Scientifique, \\ 45071 Orl\'eans, France \\ $^b$Le Studium, Orl\'eans
}

\maketitle

\begin{abstract}
The Sun provides the main energy input to the terrestrial atmosphere, and yet the impact of solar variability on long-term changes  remains a controversial issue. Direct radiative forcing is the most studied mechanism. Other much weaker mechanisms, however, can have a significant leverage, but the underlying physics is often poorly known. 

We review the main mechanisms by which solar variability may impact the terrestrial atmosphere, on time scales ranging from days to millennia. This includes radiative forcing, but also the effect of interplanetary perturbations and energetic particle fluxes, all of which are eventually driven by the solar magnetic field.

\bigskip

\noindent
slightly expanded version of an article to appear in \textit{Comptes Rendus Geoscience} (2009), special issue on \textit{The atmosphere observed from space}.

\bigskip


\end{abstract}

\section{Introduction}
\label{geosci_intro}

In two decades, the connection between solar activity and the Earth's atmosphere has moved from a mere curiosity to a hotly debated topic. Many reviews have been written, emphasising either the radiative forcing from a solar viewpoint \cite{foukal06, froehlich04, lean97, lean05, lean98b}, or from a terrestrial viewpoint \cite{haigh05, haigh07}, solar variability in general \cite{ipcc07, benestad06, calisesi07, friischristensen00, lockwood05, pap04, rind02}, historical aspects and long-term effects \cite{bard06, beer06, dejager05, hoyt97, usoskin08, versteegh05}, and other, indirect mechanisms \cite{marsh03, tinsley08}. Here we review the solar inputs to the terrestrial atmosphere and focus on their origin, the underlying physics and their observation.

The Sun-Earth connection is a world of paradoxes. Until recently, this seamless system was widely considered as a stack of independent layers, and only in recent times did the interactions between these layers really attract attention. The role of the Sun in our solar system goes undisputed, and yet the effect of solar variability on the atmosphere remains quite controversial. As we shall see later, the main mechanisms by which the Sun affects the Earth are not the most immediate ones in terms of energetic criteria. 

The Sun -- like any living star -- continuously radiates energy outward into the heliosphere. The radiated energy is carried by (i) electromagnetic waves over a frequency band ranging from radio waves to hard X-rays, (ii) a stream of hot plasma (the solar wind) consisting primarily of electrons and protons with a small fraction of heavier ions, (iii) an interplanetary magnetic field (IMF) which is carried along with the solar wind (often referred to as a frozen-in magnetic field), and (iv) violent solar outbreaks such as solar flares and coronal mass ejections (CME) \cite{kamide07}.

The solar radiative output is nearly constant in time and accounts for about 1365 W/m$^2$ at a solar distance of 1 Astronomical Unit (AU), with a solar cycle dependent variation of the order of 0.1 \%. Under quiet solar conditions the flow rates of the kinetic energy of the solar wind bulk motion and the solar wind thermal energy amount to about $5\!\cdot\!10^{-4}$ W/m$^2$ each at 1 AU, i.e., a million times less than the radiative input. The energy flow rate of the IMF is another two orders of magnitude smaller, about  $5\!\cdot\!10^{-6}$ W/m$^2$. Yet, these different mechanisms all have a distinct impact on the terrestrial atmosphere and none of them can be ruled out a priori.

Nearly 70\% of the solar radiation that arrives at the top of the Earth's atmosphere is absorbed in the atmosphere or at the Earth's surface; the rest is immediately reflected. In contrast, the efficiency of energy transfer from the solar wind into the magnetosphere is only 1--10\%, depending on the orientation of the Interplanetary Magnetic Field (IMF). 

Wave and particle emissions are not the only means by which the Sun can influence the Earth's atmosphere. The solar wind plasma, more precisely, the IMF associated with it, modifies the rate of penetration of interstellar energetic particles into the heliosphere and eventually into the atmosphere. This has led to one of the more controversial aspects of Sun-climate studies.

In this review, we first start with an illustration of solar variability on time scales from days to decades (Sec.~\ref{geosci_variability}). Section~\ref{geosci_radiative} then addresses the solar radiative output and its effects, and Sec.~\ref{geosci_diameter} the role of orbital changes. Thereafter we focus on indirect effects, the electric circuit (Sec.~\ref{geosci_electric}, including galactic cosmic rays), atmospheric convection under quiet (Sec.~\ref{geosci_Sqsystem}) and active (Sec.~\ref{geosci_activesun}) solar conditions, and the role of the coupling with upper atmospheric layers (Sec.~\ref{geosci_atmoscoupling}). A short section then addresses the prediction of future solar activity (Sec.~\ref{geosci_future}). Conclusions follow in Sec.~\ref{geosci_conclusions}. External forcings that are not related to the Sun (such as volcanic activity) and internal forcings are not addressed.


\section{Solar variability}
\label{geosci_variability}

Solar activity affects the Earth's environment on time-scales of minutes to millions of years. The shorter time-scales are of particular interest in the frame of \textit{space weather}\footnote{Space weather mostly deals with short-term impacts and forecasting of solar-activity, with a particular focus on its societal effects: impacts on space systems, navigation, communications, ground technology, etc.} \cite{schwenn06}, but will not as much be considered here. 
Long-term changes of solar and heliospheric conditions and their manifestation in the Earth's space and atmospheric environment are typically considered to be in the realm of space climate \cite{mursula07}. 
It is often believed that only slow variations (i.e. time-scales of years and above) can affect climate. This is not fully correct in the sense that the rate of occurrence of fast transients such as solar flares is modulated in time, so that all time scales eventually matter.

\begin{figure*}[!thb] 
    \begin{center}
\includegraphics[width=0.98\textwidth]{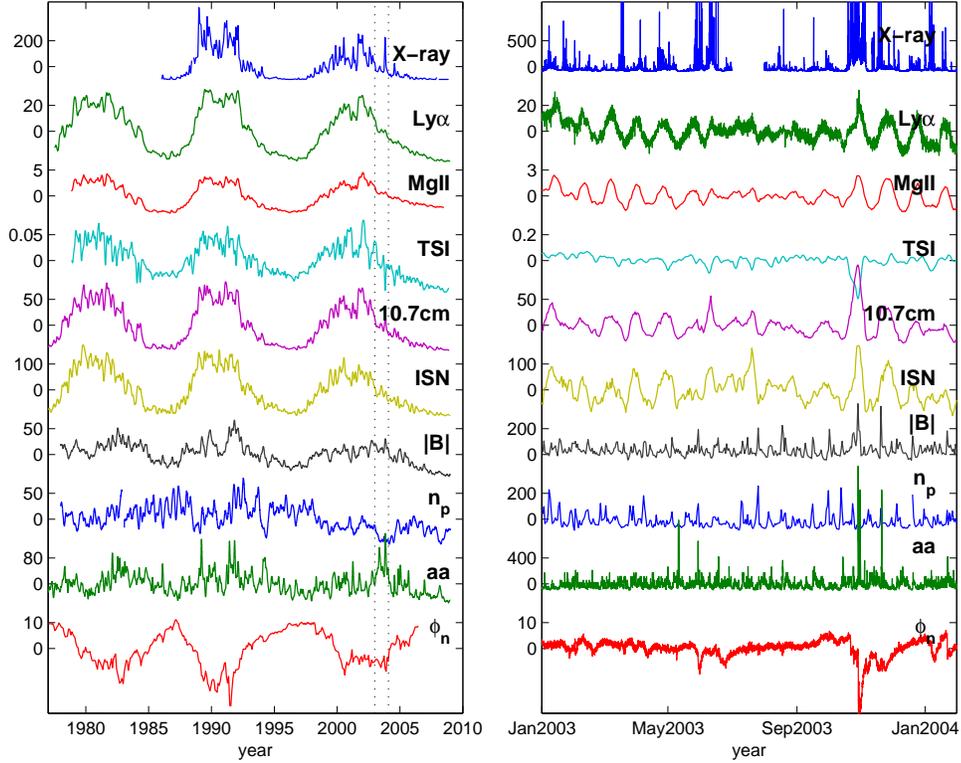} 
    \end{center}
\caption{Relative variation (in \%) of some of the key solar-terrestrial parameters. The left plots shows three decades of observations, with monthly averaged data and the right plot a one-year excerpt with hourly or daily observations. From top to bottom: Soft X-ray flux (from GOES/SEM), irradiance in the EUV (Lyman-$\alpha$ line composite from LASP, Boulder), irradiance in the UV (MgII index from NOAA), Total Solar Irradiance (TSI composite from PMOD-WRC, Davos), radio flux at 10.7 cm (from Penticton Observatory), sunspot number (ISN, from SIDC, Brussels), intensity of the magnetic field in the solar wind ($|B|$, from OMNIWeb), proton density in the solar wind ($n_p$, from OMNIWeb), $aa$ geomagnetic index (from ISGI, Paris) and neutron flux at mid-latitude ($\phi_n$, from SPIDR). All quantities are normalised with respect to their time-average. Some of the vertical scales differ between the two plots.}
\label{fig_indices} 
\end{figure*}

To give a glimpse on the complexity of solar variability, we illustrate in Fig.~\ref{fig_indices} the variation of some key solar-terrestrial parameters; several of them will be discussed in later sections. The long time interval (left panel) covers three decades only because very few accurate solar observations were available before the advent of the space age. One of the main tasks in solar-terrestrial physics today is to extrapolate these tracers backward in time.

The tracers (or \textit{proxies}, as they are usually called) of solar activity that are shown in Fig.~\ref{fig_indices}, are respectively:
\begin{itemize}
\item \textit{X-ray}: the soft X-ray flux between 0.1 and 0.8 nm, which is indicative of the energy released during solar eruptive phenomena such as flares. Most of this radiation is absorbed in the upper atmosphere (above 60 km) and above.
\item \textit{Ly$\alpha$}: the intensity of the bright H Lyman-$\alpha$ line at 121.57 nm, which is mainly emitted in the solar transition region and is absorbed in the ionosphere (above 90 km).
\item \textit{MgII}: the core-to-wing ratio of the Mg II line at 279.9 nm, which is a good proxy for the solar irradiance in the UV. This radiation is primarily absorbed in the stratosphere, where it affects ozone concentration.
\item \textit{TSI}: the Total Solar Irradiance (TSI), which represents the total radiated power measured at 1 AU, above the atmosphere. This quantity summarises the total radiative energy input to the Earth.
\item \textit{10.7 cm}: the radio flux emitted at 10.7 cm, or decimetric index. This radiation has no direct impact on climate, but it is widely used in Global Circulation Models (GCMs) as a proxy for solar activity. It is measured daily since 1947.
\item \textit{ISN}: the International Sunspot Number (ISN), one of the most ancient gauges of solar activity, with almost daily measurements since 1749.
\item $|B|$: the intensity of the interplanetary magnetic field at the L1 Lagrange point, just upstream of the Earth. 
\item $n_p$: the proton density, also measured in the solar wind. This quantity, combined with the solar wind bulk speed, gives the solar wind dynamic pressure, which is the main solar parameter to define the shape of the magnetosphere.
\item $aa$: the $aa$-index, which is a 3-hourly range measure of the level of geomagnetic field fluctuations at mid-latitudes. Its amplitude reflects the amount of magnetic energy that is released in the terrestrial environment.
\item $\phi_n$: the atmospheric neutron flux, measured on Earth, at mid-latitude. This flux is indicative of the highly energetic galactic cosmic ray flux, which is not of solar origin, but is modulated by solar activity. Part of this ionising radiation is absorbed in the middle atmosphere, where it might affect cloud condensation.
\end{itemize}

The left panel reveals a conspicuous modulation of about 11 years, which is known as the solar cycle and whose origin is rooted in the solar magnetic dynamo \cite{charbonneau05b}. Solar magnetism is indeed the ultimate  driver behind all the quantities we shall encounter here \cite{dejager05}. Its great complexity, and the wide range of spatial and temporal scales covered by its dynamics allows for a rich variety of manifestations. 

The solar cycle, which is best evidenced by the number of dark sunspots occurring on the solar surface, is probably the best documented manifestation of solar activity on our terrestrial environment. Statistically robust signatures of the solar cycle have been reported in a large variety of atmospheric records, including stratospheric temperatures \cite{labitzke88}, ozone concentration \cite{haigh94, shindell99}, changes in circulation in the middle \cite{labitzke05} and lower \cite{wilcox74} atmosphere, tropospheric temperatures \cite{crooks05},  ocean surface temperature \cite{reid91,white97}, and many more. For reviews, see \cite{haigh07, harrison99, hoyt97, versteegh05}.
 
The important point in Fig.~\ref{fig_indices} is the occurrence of the same 11-year cycle in all solar-terrestrial parameters. As a consequence, disentangling their individual impacts on the atmosphere is almost impossible without the contribution of physical models. All quantities are correlated, but not all are necessarily causally related to atmospheric changes.

A look at shorter time scales (right panel in Fig.~\ref{fig_indices}) reveals a different and in some sense much more complex picture. Some quantities exhibit an occasional 27-day modulation associated with solar rotation, but correlations are not systematic anymore. For the same reason, the properties of the 11-year cycle may not be readily extrapolated to longer time scales either.
 
Another distinctive feature of Fig.~\ref{fig_indices} is the highly intermittent nature of some quantities such as the soft X-ray flux and geomagnetic indices. The presence of rare but extreme events suggests that the rate of occurrence of such events may affect climate, even though the lifetime of each individual event is orders of magnitude below the characteristic response time of the atmosphere.


\section{The solar radiative output}
\label{geosci_radiative}

The largest solar energy input to the terrestrial environment comes through electromagnetic waves. The Sun radiates over the entire spectrum, with a peak in the visible part (400-750 nm). The actual shape of the spectrum is dictated by the composition of the solar atmosphere and its temperature, which increases from near 6000 K in the photosphere to millions of degrees in the corona.

The bulk of the solar spectrum is relatively well described by the emission of a black-body at 5770 K. On top of this smooth spectrum come numerous discrete features associated with absorption and emission processes \cite{lean97}. The ultraviolet part of the solar spectrum (UV, 120-400 nm) is partly depleted by such absorption processes, whereas the Extreme-UV (EUV, 10-120 nm) is strongly enhanced by contributions from the hotter part of the solar atmosphere. The visible and near-infrared contributions both represent about 45 \% of the total radiated power, whereas the UV represents about 8\% and the EUV less than 10$^{-3}$ \%. Although the different layers of the solar atmosphere are strongly coupled by the solar magnetic field, the variability of the solar spectrum is remarkably complex and cannot properly be described by a single parameter.


\subsection{The total solar irradiance}
\label{geosci_tsi}

When studying the Earth's global energy budget (see \cite{kiehl97} and also the chapter by R. Kandel in this volume), the solar radiative forcing is often represented by a single convenient parameter, called  \textit{total solar irradiance} (TSI). The TSI is the power integrated over the entire solar spectrum. For a long time, it was believed to be constant, hence its ancient name \textit{solar constant}. 

The TSI can only be measured from space since the terrestrial atmosphere absorbs part of the radiation. The first measurements started in 1978 and revealed a small but significant variation. Several missions have measured the TSI since, giving an average value of 1365 W/m$^2$ \cite{froehlich06}. The relative amplitude\footnote{defined here as (maximum-minimum)/time average.} over a solar cycle is 0.1 \% but short-term variations of up to 0.25 \% may occur during periods of intense solar activity \cite{woods04}. 

\begin{figure}[!htb] 
    \begin{center}
\includegraphics[width=8cm]{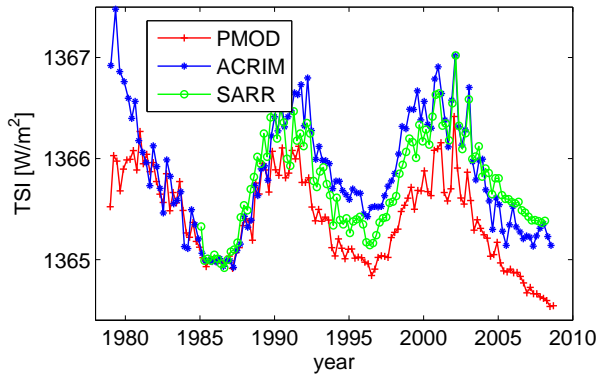} 
    \end{center}
\caption{Comparison of three composites of the total solar irradiance, averaged over 81 days. The composites are: PMOD version d41-61-0807  \cite{froehlich06}, ACRIM version 11/08 \cite{willson03}, and SARR version 3/08 \cite{mekaoui08}. For better visibility, all curves have been shifted vertically to share the same average value for 1986-1987.
} 
\label{fig_3tsi} 
\end{figure}

Different TSI observations agree on the short-term relative variability, but significant differences exist between their long-term trends. There exist today three composites of the TSI, based on how the data from different instruments are stitched together, see Fig.~\ref{fig_3tsi}. The disagreement between these three versions regarding the existence of a secular trend has fuelled a fierce debate. Indeed, the composite of the PMOD group \cite{froehlich06} suggests the existence of a recent downward trend in the TSI, whereas the ACRIM group \cite{willson03} claims the opposite.

Two key issues with the TSI are the origin of its variability and the reconstruction of past values. The Sun is photometrically quiet and the short-term variability mainly results from a competition between an irradiance deficit due to sunspots and an enhancement due to bright photospheric features called faculae \cite{froehlich04}. The two effects are connected, but the variability affects different spectral bands. The secular trend in the TSI is more directly related to weak changes in brightness during spotless periods (called \textit{quiet Sun}), which means that trends are best observed by comparing minima in the solar cycle. The origin of these slow brightness changes is still unclear, although it is certainly related to the solar magnetic field \cite{foukal06}.

A reconstruction of pre-1978 values of the TSI is of course a major issue for climate studies. There is strong observational evidence for solar surface magnetism to be the major driver of TSI changes on time scales of days to years \cite{krivova03}. Based on this, Fligge et al. \cite{fligge00} developed a semi-empirical model for reconstructing TSI changes from the surface distribution of the solar magnetic field, using solar magnetograms inferred from solar images of the Ca K line emission. Unfortunately, few images exist before 1915, which limits the applicability of the method. 

The only direct solar proxy that is sufficiently homogeneous for reconstructing the TSI back to the Maunder minimum is the sunspot number. The Maunder minimum (1645-1715) is of particular interest since the Sun was very inactive at that time and temperatures in the Northern hemisphere were unusually low \cite{eddy76, shindell01}. By using reconstructions of the sunspot number going back to 1610 as inputs to open magnetic flux transport simulations, several authors \cite{krivova07,lean00b, wang05} have demonstrated that the TSI was lower during the Maunder minimum than today. The uncertainty on the actual change in TSI, however, is high. Present estimates give a change in radiative forcing (the net downward radiative flux) from +0.06 to +0.3 W m$^{-2}$ \cite{ipcc07b}, which is equivalent to a $\Delta T = $+0.04 to +0.18 K increase in global temperature since the Maunder minimum. The Intergovernmental Panel on Climate Change (IPCC) concludes that this bare change is insufficient to explain the observed global temperature increase \cite{ipcc07b}. The same conclusions hold for reconstructions made since 1978.

For TSI reconstructions on time scales of centuries to millennia, a different approach must be used. The most reliable proxies are cosmogenic isotopes such as $^{14}$C and $^{10}$B, whose production rate  is modulated by solar activity \cite{bard06}. Bard et al. \cite{bard00} have shown that relative variations in the abundance of such cosmogenic isotopes are in excellent agreement with sunspot-based TSI reconstructions. There have been attempts to reconstruct solar activity up to hundreds of thousand years in the past \cite{usoskin08}. For such long periods, however, the slowly but erratically varying geomagnetic field becomes a major source of uncertainty. Discrepancies between paleomagnetic reconstructions based on different deep-sea cores today are still too important to properly quantify the solar contribution 20 kyr and more backward \cite{bard06}. 

The relatively small impact of solar radiative forcing on climate has been questioned by several. Scafetta and West, for example, used a phenomenological model to conclude that at least 50\% of the global warming observed since 1900 had a solar origin \cite{scafetta05b, scafetta07}. Three recurrent arguments are: (i) recent solar activity is better reflected by the TSI composite from the ACRIM group than from the PMOD group; (ii) short-term statistical fluctuations and longer-term cycles have distinct effects \cite{scafetta05b}, which may explain why such clear signatures of solar cycles (11-year, but also the weaker 90-year Gleissberg cycle) have been found in atmospheric records; (iii) feedback mechanisms are not sufficiently well understood and positive feedback may be much stronger than expected \cite{pap04,stott03}. Lockwood and Fr\"ohlich \cite{lockwood07} argue that the PMOD composite is the most reliable, and so solar activity has not increased at the end of the 20th century. Objections against (ii) and (iii) have been made by climate modellers who do not see evidence for such effects in GCMs, see for example the comment by Lean \cite{lean06b}.  

Most of the TSI consists of visible and near-infrared radiation, which are primarily absorbed by oceans and land surfaces, and in the lower troposphere by water vapour and by CO$_2$. For that reason, a direct connection between TSI change and tropospheric temperature change can be established. This direct forcing is insufficient to explain the observed temperature increase. However, several effects such as the hydrological cycle \cite{shindell06} and  stratospheric water vapour feedback \cite{stuber01} could have an impact on the forcing-response relationship. The debate continues unabated.


\subsection{The solar spectral irradiance}
\label{geosci_spectral}

A significant portion of the solar radiative output does not account for a direct radiative forcing because it is absorbed in the middle and upper atmosphere where it affects photochemistry. Spectrally resolved observations are required to study these effects. 

\begin{figure}[!htb] 
    \begin{center}
\includegraphics[width=8cm]{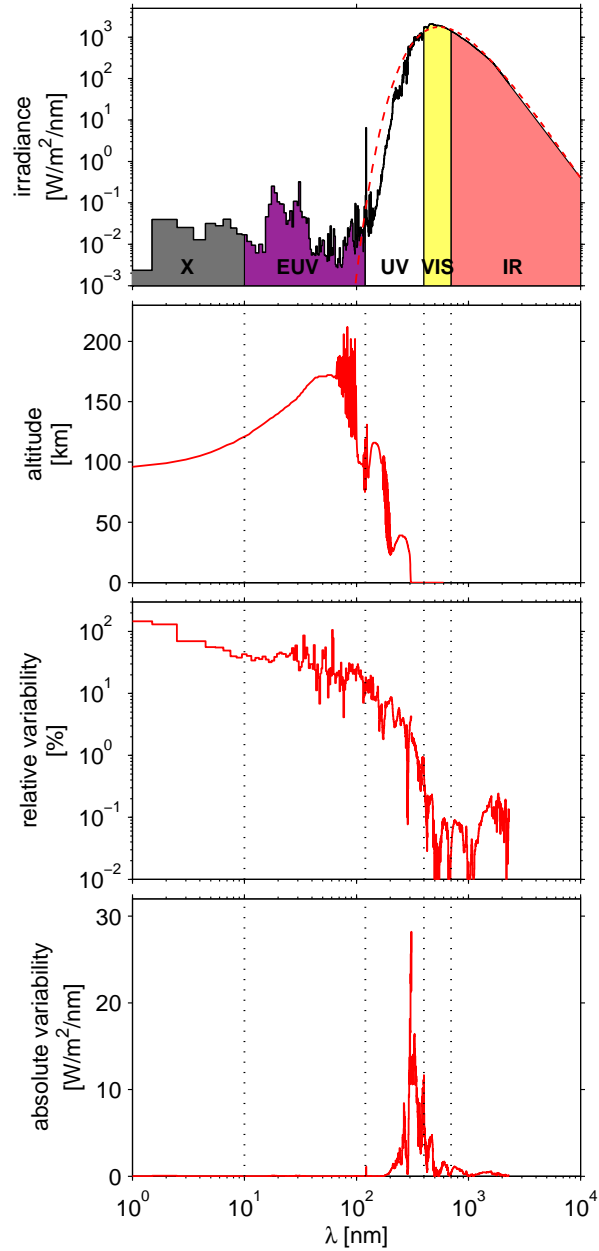} 
    \end{center}
\caption{From top to bottom: the solar irradiance spectrum with (dashed) a black-body model at 5770 K, the altitude at which the UV and EUV components are predominantly absorbed (unit optical depth), the relative and absolute variability of the irradiance from solar maximum to solar minimum. All these results refer to the [Oct. 2003--Jan. 2009] time span. This plot is based on observations from SORCE/XPS, TIMED/EGS, SORCE/SOLSTICE and SORCE/SIM.
} 
\label{fig_sorce_spectra} 
\end{figure}

The principal features of the solar spectrum and its variability are illustrated in Fig.~\ref{fig_sorce_spectra}. The main result is the large relative variability in the UV band and below, which exceeds that of the TSI by orders of magnitude. In absolute terms, this spectral variability peaks in the UV between 200 and 400 nm. Below 310 nm, this radiation is strongly absorbed in the mesosphere (from 50 km to about 80-90 km), and in the stratosphere by the ozone Hartley band (see the chapter by S. Godin-Beekman in this volume). During periods of intense solar activity, the ozone concentration thus increases, heating the stratosphere and higher layers, which affects the downward radiative flux. This also impacts the meridional temperature gradient, altering planetary and gravity waves, and finally affecting global circulation \cite{haigh94}. Haigh first introduced this general picture, which is now widely accepted \cite{haigh07, larkin00, shindell01}. The main effects are a warming of the upper and lower stratosphere at low and middle latitudes, and a strengthening of the winter stratospheric polar night jet. Direct heating by absorption of the UV can explain most of the temperature response in the upper stratosphere  but not in the troposphere and lower stratosphere. The final temperature response depends critically on the ozone concentration profiles and on details of the coupling mechanisms. These mechanisms are non-linear, and so a meaningful radiation budget cannot be established without resorting to GCMs. These models show important discrepancies and yet, recent comparisons seem to converge toward a mean model response of up to about 2.5 \% in ozone and 0.8 K in temperature during a typical solar cycle \cite{austin08}.

Less than 0.01 \% of the total irradiance comes from wavelengths below 200 nm. This small contribution is mostly absorbed in the lowermost ionosphere, where photodissociation affects the local composition and generates heat. Because this part of the solar spectrum is highly variable, it has a noticeable effect. On time scales of hours to days, solar flares, for example, can increase the electron density by orders of magnitude \cite{lilensten08}. Long-term signatures of solar activity are also evident in many ionospheric parameters; the most conspicuous one is the 11-year solar cycle \cite{jarvis06, lastovicka09}. The solar-cycle dependence of the height of constant plasma density in the lower 
ionosphere is attributed to the competing effects of a higher ionisation rate (resulting in higher plasma density at a given fixed height) and increased atmospheric heating and upwelling (resulting in lower plasma density at the same height) at solar maximum as compared to solar minimum. A slow global cooling has also been observed \cite{bencze07}, similar to that found in the meso- and stratosphere. This global cooling is most likely related to a contraction of the atmosphere due to an increasing concentration in greenhouse gases.


We conclude at this stage that the photochemical and dynamical impacts of the solar UV component have a significant leverage on the stratosphere and on climate. According to the IPCC \cite{ipcc07b}, this mechanism cannot explain the temperature increase observed during the 20th century; it would require an amplification that is not reproduced by present GCMs. Three important issues are: (i) to better understand the physical coupling mechanisms within the middle atmosphere and with the lower atmosphere; (ii) to include in GCMs which started in the lower atmosphere a proper description of the often overlooked upper atmosphere and in originally thermospheric CGMs a proper link to the lower atmosphere, and (iii) to improve the solar inputs to these models in order to obtain a better response of ozone concentration versus time and position.

Concerning the last issue, we note that solar spectral irradiance observations are highly fragmented and inaccurate. Indeed, such measurements must be carried out from space, where detectors suffer from degradation.  An ``overlap strategy'' is frequently used, where successive satellite experiments are directly compared to improve their long-term accuracy. For the TSI, uncertainties of 1 part in 10$^5$ per annum can be obtained, whereas for the EUV-UV range, errors of more than 50 \% unfortunately are not exceptional. 

The first continuous observations of the EUV-UV spectrum started in 2002 with the TIMED mission \cite{woods05}, later complemented by SORCE. Because of this severe lack of radiometrically accurate observations, most users of UV data, including climate modellers, have resigned to using proxies. The radio flux at 10.7 cm (or f10.7 index, see Fig.~\ref{fig_indices}) is often used in atmospheric studies, for it can be conveniently measured from ground. The MgII index \cite{thuillier01} has been advocated as a better proxy for the UV, but none of these quantities can properly reproduce the spectral variability \cite{ddw09}.


\section{Orbital changes and solar diameter variations}
\label{geosci_diameter}

Orbital changes, and variations in the solar diameter have very little in common. Both, however, lead to a slow modulation of the solar irradiance that can be described in geometrical terms. In this sense, they fall under the preceding section. Orbital changes are well understood \cite{crucifix06} and are discussed in the chapter by D. Paillard (this volume).

The evidence for a variability of the apparent solar diameter has on the contrary remained elusive. Ground and space observations yield relative amplitudes of less than $0.06 \%$ over one cycle but do not agree \cite{thuillier05}. The effect on climate is likely to be small, but cannot be ruled out. The upcoming Picard mission, which will be launched in 2010, precisely aims at measuring the solar diameter during the rising phase of the solar cycle with unprecedented accuracy.


\section{Solar impact on atmospheric electricity}
\label{geosci_electric}

Atmospheric electricity is an old field of research but its role in the Sun--Earth coupling has recently attracted considerable interest and controversy. The effect of ions on the atmosphere is discussed in more detail by E. Blanc (this volume); here we concentrate on the role of the Sun only.


\subsection{Effect of the atmospheric current}

A minute current of $\sim2$ pA/m$^2$ permanently flows down from the ionosphere through the troposphere to the terrestrial surface, generating charges that are capable of affecting the nucleation of water droplets to form clouds. This current responds to internal but also to solar forcings, providing a mechanism by which solar activity affects various atmospheric parameters such as cloud cover, temperature and precipitation \cite{rycroft00, tinsley07}. Tinsley \cite{tinsley08} has shown that there are at least four indirect solar inputs which modulate the process: (i) variations in the galactic cosmic ray flux, mediated by solar activity (see Sec.~\ref{geosci_cosmics}); (ii) solar energetic particle fluxes that are occasionally generated by intense solar flares or CME associated shocks; (iii) relativistic electrons coming from the Earth's radiation belts and (iv) polar cap ionospheric electric potential changes (see Sec.~\ref{geosci_Sqsystem}). The latter two are mainly induced by geomagnetic activity driven by interplanetary perturbations.

Most of the mechanisms listed above occur erratically and on time scales of days and so their long-term impact is difficult to assess. Recent advances have been made in the study of transient luminous events (see the chapter by E. Blanc in this volume), which provide an unexpected energy link between the lower ionosphere and the upper troposphere. 


\subsection{Effect of galactic cosmic rays}
\label{geosci_cosmics}

During the active part of the 11-year solar cycle the solar magnetic field and its heliospheric extension, the IMF, are generally stronger and more turbulent than around solar minimum. A stronger IMF will more successfully guide and deflect interstellar protons than a weaker IMF, with the result that the solar cycle imposes an 11-year modulation on the flux of galactic cosmic rays (GCRs) reaching the Earth's atmosphere. 
The contribution of cosmic rays to ion production in the atmosphere on short and long time scales is well established, see for instance the review by Bazilevskaya et al. \cite{bazilevskaya08}. At present at least three models in use describe this process: one developed in Oulu \cite{usoskin06c}, another in Bern \cite{desorgher05} and a third one in Sofia \cite{velinov01, velinov07}. A comparison of model simulations with balloon-borne ion density measurements has shown that models and measurements are in good agreement \cite{usoskin09}.

Svensmark and co-workers \cite{svensmark07, svensmark97} promoted a mechanism in which an increased intensity of the GCR flux is, at least in part, responsible for an enhanced density of free ions and electrons in the troposphere. The free electrons, liberated by cosmic rays, assist in producing ionised aerosols which in turn should act as water vapour condensation nuclei in the troposphere. Tinsley and co-workers \cite{tinsley07} suggested that a GCR flux modulation changes the aerosol ionisation which in turn changes the ice nucleation efficiency of the aerosol. 
In both cases, the net effect is an enhancement of the global low-altitude cloud coverage, a modification of the Earth albedo and eventually a modulation of the global tropospheric temperature in correlation with the 11-year solar activity cycle. In short, it is suggested (e.g. \cite{svensmark07}) that the cloud coverage is modulated by the solar cycle, at least at heights below some three kilometres.

This view is cautioned by others. Sun and Bradley \cite{sun02} cast doubt on the usefulness of the selection of data used by Svensmark and Friis-Christensen \cite{svensmark97} and demonstrate that results become different if different analysis intervals are considered. They conclude that no solid observational evidence exists for the suggested GCR--cloud coverage relation. Harrison and Carslaw \cite{harrison03} and Usoskin \cite{usoskin08} conclude that neither the GCR--cloud coverage link proposed by Tinsley nor the one proposed by Svensmark can be excluded but find that some elements in the chains of both mechanisms remain contentious, and they doubt whether the processes are efficient enough to contribute significantly to a modulation of low cloud formation. Sloan and Wolfendale \cite{sloan08} estimate that on a solar cycle scale, less than 23\% of the 11-year cycle change in the globally averaged cloud cover is due to the change in the rate of ionisation from the solar modulation of cosmic rays.

The controversy is still going on, and the lack of accurate long-term observations of cosmic ray intensity and especially global cloud coverage presently does not allow to accept or discard a potential influence of the GCR-cloud connection on long-term changes of the tropospheric mean temperature. 
The CLOUD experiment that is planned at CERN should help better quantify the cloud formation rate \cite{svensmark07}.
Experimental evidence gathered so far appears to suggest that on short time scales (a few days) and on interannual time scales a link between cosmic ray flux and low cloud coverage exists. The correlation between low cloud area coverage and cosmic ray induced ionisation has been found to be dependent on latitude and geographic region.
It is significantly positive at mid-latitudes but poor (and possibly negative) in the tropics \cite{usoskin04b, palle04}. Depending on the time interval considered better correlations exists over the Atlantic (1983-2000) or over the Pacific (1983-1993) \cite{palle04}. Europe and the North and South Atlantic exhibit the best correlation over the period 1984-2004 \cite{usoskin08b}. The pronounced regional variation of the correlation eventually results in a poor global correlation \cite{usoskin04b}.

Let us stress again  that all solar variability is eventually driven by the solar magnetic field, and so it is difficult to quantify the real contribution of each mechanism. As an illustration, Lockwood et al. \cite{lockwood99d} found the open solar magnetic flux to increase during the 20th century. This results in an increased shielding against GCRs and possibly a reduced cloud coverage. The same open magnetic flux, however, is also strongly correlated with the TSI \cite{lockwood01} and with the level of geomagnetic activity, both of which lead to a temperature change.


\section{Atmospheric convection under quiet solar conditions} 
\label{geosci_Sqsystem} 

Under quiet solar conditions the transfer of energy from the Sun into the Earth's 
atmosphere leads to the development of an electric current system (the solar quiet 
or \Sq system) which consists of two components, one driven by solar electromagnetic 
radiation (\Sqq) and the other by the interaction between the solar wind and the geomagnetic 
field. Note that the influence of the geomagnetic field on the motion of charged 
particle is rather strong such that the electrons (which above some 70-80 km altitude are 
little affected by collisions and are the more important carriers of ionospheric 
electric currents) move preferentially perpendicular to both the electric and 
magnetic fields (known as Hall effect). 

Solar UV/EUV heating increases the scale height of the neutral constituents and 
causes their daytime upwelling, which is accompanied by a systematic neutral gas 
redistribution via tidal winds. The ionised part of the upper atmosphere between 
about 90 and 140 km altitude, dynamically strongly coupled to the neutral gas 
via collisions between ions and neutral atoms and molecules, expands and contracts 
with the neutral gas. As this motion takes place in the presence of the geomagnetic 
field the charged particles experience a dynamo force and move along closed stream 
lines. They form the \Sqq current system, which is significant between northern 
and southern auroral latitudes but practically negligible at polar cap latitudes. 
The corotation electric field (due to the frictional coupling of the neutral 
atmosphere to the Earth rotation) exercises a strong influence at low, middle and
subauroral latitudes and imposes a systematic eastward shift on the \Sqq \ pattern.   

Seen from an observer at a fixed point in a Sun-Earth coordinate system, 
i.e., not rotating with the Earth (for instance, at rest in a geocentric solar 
magnetospheric [GSM] system), the solar wind together with the IMF create a 
$\vec{v}_{SW}\!\times\!\vec{B}_{IMF}\,$ electric field, usually termed 
``solar wind merging electric field'' along the high-latitude magnetospheric 
boundary (with $\vec{v}_{SW}$ and $\vec{B}_{IMF}$ denoting the solar wind 
bulk speed and IMF vectors, respectively). 
The electric field maps down to the Earth's atmosphere along geomagnetic field 
lines (which can be considered equipotential lines in the magnetosphere) and 
is observed as an electric field from dawn to dusk across the polar cap. 
This electric field, combined with the geomagnetic field (downward in the 
northern and upward in the southern polar cap) supports a Hall current from 
the nightside to the dayside across the polar cap, closed by return currents 
(known as auroral electrojets) at slightly lower but still auroral latitudes. Such return currents must flow in the ionosphere because the ionospheric Hall currents are divergence free. This is the second contribution to the \Sq currents. The coupling of the atmosphere to the rotating Earth and the magnitude of the east-west component of the IMF modify the preferential orientation of the 
convection pattern in the sense that it may become more or less shifted, 
mostly in westward but sometimes in eastward direction.  

Although the rate of solar radiation on the topside atmosphere depends solely on geographic latitude and longitude the \Sq current system also depends on geomagnetic latitude and longitude, as a result of the ionospheric plasma density distribution. The latter is not only governed by charge production via UV and EUV radiation but also by the electric conductivity tensor, which depends on the orientation of the geomagnetic field vector. For instance, close to the geomagnetic equator the magnetic field is nearly horizontal. The only way to move electric charges across the geomagnetic field is along the equator as any vertical electric current would immediately be quenched by space charges accumulating at the lower and upper boundaries of the ionosphere. This effect facilitates considerably the establishment of a narrow electric current strip in the dayside upper atmosphere along the geomagnetic equator (known as equatorial electrojet). 

The \Sq current system is strongly dependent on season, with a remarkable 
increase in the summer and a decrease in the winter hemisphere. The \Sq system 
further depends on the solar cycle; the somewhat higher average solar wind speed 
and the enhanced atmospheric ionisation due to more intense UV/EUV radiation 
and energetic particle precipitation increase the electrical conductivity and 
contribute to more intense ionospheric electric currents during the maximum 
and early declining phases of the solar cycle. 

\begin{figure*}[ht] 
\vspace*{2mm} 
\centering 
\includegraphics[width=0.8\textwidth]{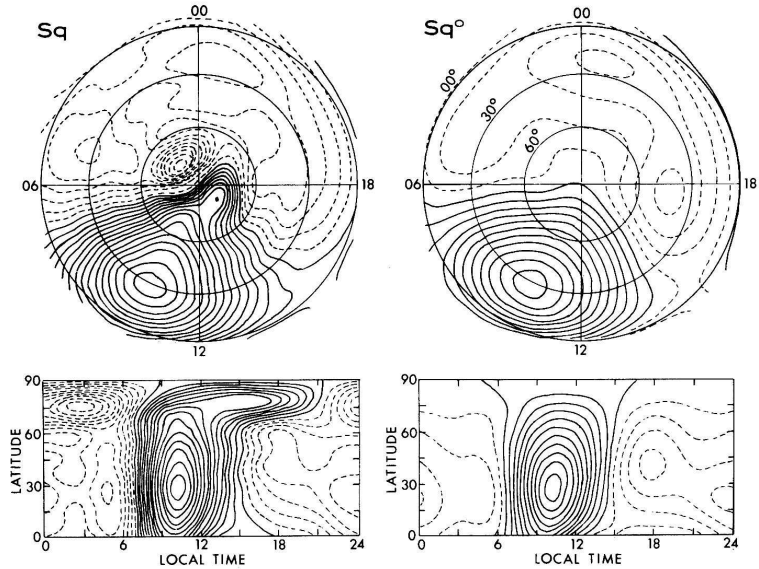} 
\caption{Ionospheric \Sq current system as inferred from ground-based magnetometer observations at 40 sites over the period May-June 1965. Right hand side: tidal currents only, left-hand side: combined tidal and polar cap currents. Current intensity between adjacent lines is 10 kA counterclockwise (solid) and clockwise (dotted) (from \cite{matsushita75}).
} 
\label{fig:Sq_system} 
\end{figure*}

Figure \ref{fig:Sq_system} (from \cite{matsushita75}) 
shows the \Sq current system generated by solar electromagnetic radiation alone (\Sqq, right hand side) and the combined electromagnetic and solar wind generated \Sq system (left hand side). 

\section{The impact of solar activity on the Earth's atmosphere}
\label{geosci_activesun}

The steady-state conditions representing the quiet Sun are not typical for the 
maximum and early declining phases of the solar cycle. The impact of short-term 
(transient) events on the Earth's atmosphere can be profound \cite{kamide07}. Several types of 
eruptions are known to occur, with solar flares and coronal mass ejections (CMEs) 
being the most violent ones (as far as the effects on the Earth's environment are 
concerned). Just as under quiet solar conditions both electromagnetic radiation 
and solar energetic particle fluxes play important roles for the state of the 
upper atmosphere under the various types of active solar conditions. 

Solar flares, a bursty type of energy release, radiate broad-band electromagnetic 
waves whose intensities are much higher than steady-state solar radiation. The rather 
strong X-ray component associated with flares penetrates deep into the atmosphere  
and enhances the ionisation level between 60 km and 90 km altitude. This has deleterious effects on HF radio wave propagation. 

Some solar flares and CMEs are accompanied by streams of very energetic protons 
(up to hundreds of MeV) ejected from the Sun and accelerated in the solar corona and beyond. 
Unlike the typical solar wind protons ($\approx$ 1 keV) these high-energy protons 
can penetrate into the outer magnetosphere nearly unhindered by the geomagnetic 
field (which normally shields the Earth environment from the direct entry of 
solar wind particles) and propagate along the field lines toward the Earth. 
Protons with energies up to 10 MeV ionise the polar atmosphere at altitudes 
significantly below 100 km, which facilitates considerably the absorption of 
HF radio waves propagating at polar latitudes (referred to as PCA -- polar cap 
absorption). 
The flare-associated proton flux may last for several days which is the time 
it takes to bring the plasma density back to a normal level. 

A different category of solar activity, with less profound effects on the average, 
follows a recurrent pattern. At the boundary between low speed ($\approx$ 400 km/s) 
and high speed ($\approx$ 700 km/s) solar wind flow regimes one often observes a 
shock front that is produced by the high speed plasma pushing the low speed plasma. 
The flow regime boundary is fixed to the solar surface, rotates with the Sun and 
is likely to persist for longer than one solar rotation such that the associated 
solar wind structures show a tendency to hit the Earth's space environment again 
after one solar rotation (approximately 27 days).

\begin{figure*}[ht]
\centering 
\includegraphics[height=0.98\textwidth,angle=-90]{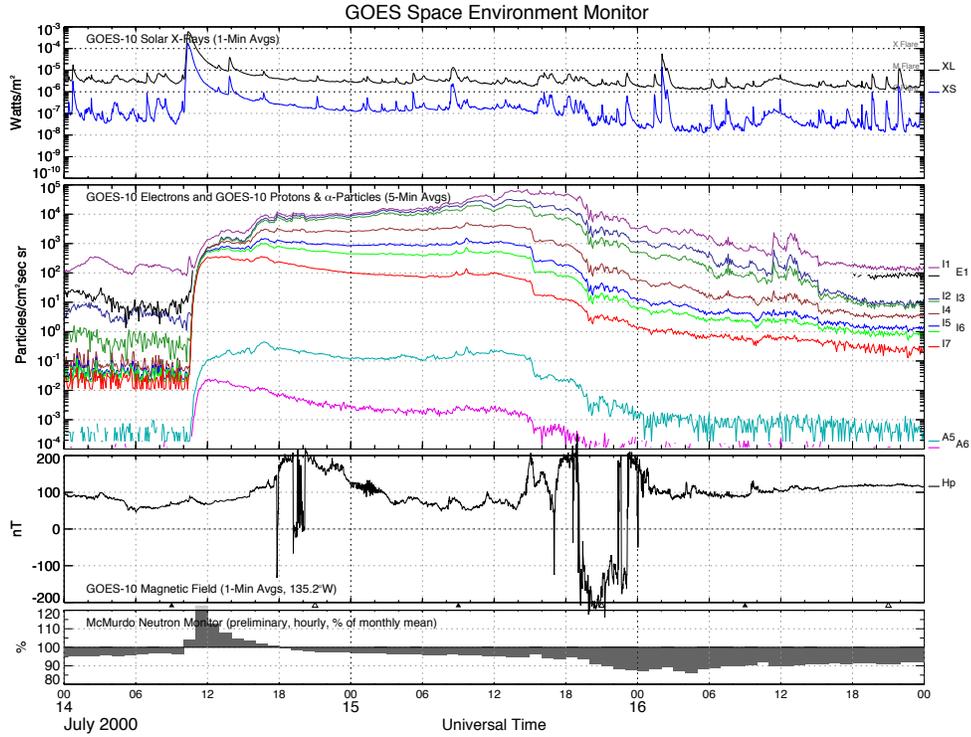} 
\caption{Observations by the SEM instrument onboard GOES-10 during the Bastille day storm. Top panel: soft X-ray intensity in the 0.05-0.3 nm (XL) and 0.1-0.8 nm (XS) bands.  Second panel: Energetic particle flux intensity. Protons $>$1 MeV (I1), $>$5 MeV (I2),  $>$10 MeV (I3), $>$30 MeV (I4), $>$50 MeV (I5), $>$60 MeV (I6), $>$100 MeV (I7),  electrons $>$2 MeV  (E1 -- mostly no data), $\alpha$ particles 150-250 MeV (A5),  300-500 MeV (A6). Third panel: magnetic field perpendicular to GOES orbital plane (i.e., practically in geographic northward direction).
 Fourth panel: neutron flux from the McMurdo neutron monitor in the Antarctic. Note the ground level enhancement  at 11 UT on 14 July and the Forbush decrease (deepest in early morning of 16 July).  Open and closed triangles mark the position of GOES-10 at local noon and  midnight,  respectively. [Figure generated by NOAA's National Geophysical Data Center]}
\label{fig:GOES_Bastille} 
\end{figure*}

Figure \ref{fig:GOES_Bastille} (from NOAA-NGDC) shows, among other parameters, 
solar X-ray and energetic particle fluxes observed at geostationary orbit during 
the geomagnetic storm on 14 July 2000 which became famous as the "Bastille day storm". 
On 14 July the X-ray fluxes in both channels reach X-class intensity which is 
considered severe by space scientists. While the X-ray flux returns to near 
pre-flare intensities after several hours the particle flux remains highly 
elevated for more than a day and moderately elevated for several days. 

Solar energetic particles can penetrate the Earth's atmosphere down to stratospheric 
and even tropospheric heights. For instance, chemically induced changes in the abundance 
of nitric oxide constituents in the stratosphere resulting from such fluxes were 
observed with the UARS satellite \cite{jackman06}. In another case extremely energetic 
solar cosmic rays associated with the intense solar X-ray flare and CME of 20 January 
2005 led to a substantial ground level enhancement and an increase of the aerosol 
density over Antarctica as inferred from the TOMS Aerosol Index \cite{mironova08}.

Auroral activity, triggered by the impact of solar activity on the Earth's 
magnetosphere, is one of the various sources of atmospheric gravity waves. 
Gravity waves play a significant role in the momentum and energy budget of 
the mesosphere and lower thermosphere \cite{fritts03}. 

Both electromagnetic radiation and charged particle precipitation into the atmosphere 
can lead to a modification of the neutral air density in the upper atmosphere.
Excessive UV and EUV radiation associated with solar activity, and to a smaller 
extent keV particle precipitation and Joule heating (caused by the motion of the 
ionospheric plasma forced by strong electric fields) can heat the atmosphere at the 
altitudes of Low Earth Orbiting (LEO) satellites -- between about 300 km and more 
than 1000 km above the ground -- thereby increasing the neutral air density at
a given height and eventually leading to increased satellite drag. At the lowest 
satellite altitudes (300-400 km) the air density can reach several times the 
value typical for quiet conditions.  

A connection between solar activity and the atmosphere that is specific 
to the Antarctic continent was proposed by Troshichev \cite{troshichev08}.
The solar wind merging electric field maps, via field-aligned currents, 
down to the atmosphere to establish a trans-polar cap electric potential 
whose changes can, via electric connection to the troposphere, influence 
the large-scale vertical circulation system that forms above the Antarctic 
continent in the winter season. 
In this circulation system air masses descend above the central Antarctic 
ridge and ascend near the coast. If the vertical winds become very strong 
(for instance, as a result of field line merging at the magnetopause) 
they disturb the thermal equilibrium which results in an increased cloud 
coverage over Antarctica, and they disturb the large-scale horizontal wind 
system, thereby quenching the circumpolar wind vortex. 
Indirect evidence for this effect was inferred from regular meteorological 
observations made at Antarctic stations.


\section{Coupling of atmospheric layers}
\label{geosci_atmoscoupling}

The coupling between the ionised and neutral gas components of the upper atmosphere 
up to about 140 km is a two-way process. If electric field and neutral wind measured 
in an Earth-fixed reference frame are denoted by $\vec{E}$ and $\vec{u}$, respectively, 
the electric current density in the presence of the geomagnetic field, $\vec{B}_0$, is 
expressed as $\vec{J}\!=\!\mathbf{\Sigma}\,(\vec{E}\!+\!\vec{u}\!\times\!\vec{B}_0)$ 
with $\mathbf{\Sigma}$ denoting the electric conductivity tensor. 
An electric field (of external origin, for instance) influences the ion velocity and, 
via collisional coupling, the neutral gas while the neutral wind (due to pressure, 
gravity and the Coriolis force, for instance) is equivalent to a 
$\vec{u}\!\times\!\vec{B}_0$ electric field (in an Earth-fixed frame) and influences 
in return the ion and electron velocities. In other words, solar energy may be 
transferred from the electrically charged to the neutral component of the upper 
atmosphere via frictional heating while kinetic energy may be transferred from 
the neutral to the charged component via a neutral wind associated electric field.

In addition to dynamic coupling between the neutral and electrically charged components of the ionosphere it has become evident that different atmospheric height regions are also coupled. Planetary waves are prime candidates for linking different altitudes \cite{salby84}. They are large-scale oscillations of the lower, middle and upper atmosphere with periods preferentially (but not exclusively) near 5, 10 and 16 days. In some cases planetary waves are generated in the lower atmosphere (troposphere and stratosphere) and propagate upward into the middle and upper atmosphere. In other cases they appear to have been generated in the middle atmosphere and propagate latitudinally. 

Goncharenko and Zhang \cite{goncharenko08} conclude that seasonal trend, solar flux and geomagnetic 
activity cannot account for temperature variations in the thermosphere which they had 
observed during an incoherent scatter radar campaign in Jan-Feb 2008. They suggest 
that the variations are associated with stratospheric warming and hence demonstrate 
a link between the lower and the upper atmosphere. Yi{\u g}it et al. \cite{yigit08} demonstrate 
the penetration of gravity waves and subsequent momentum deposition from the lower 
troposphere and stratosphere to the middle thermosphere. 

Supported by the observational evidence acquired over the years it became clear that kinetic and electromagnetic coupling between atmospheric layers exists, and the need for developing coupled atmosphere-thermosphere-ionosphere-plasmasphere models emerged. As a consequence, global circulation models (GCMs) of the terrestrial upper atmosphere have evolved. About a decade ago the time-dependent 3-dimensional Coupled Thermosphere Ionosphere Plasmasphere (CTIP) model was developed \cite{millward96}. The CTIP model consists of three distinct components, a global thermosphere model, a high-latitude ionosphere model and a mid- and low-latitude ionosphere/plasmasphere model. 

The Coupled Middle Atmosphere and Thermosphere model (CMAT) is one of the advanced models ultimately derived from the CTIP model. Its range of validity was originally extended down to 30 km altitude \cite{harris01}, and a further improved version (CMAT2, \cite{cnossen08}) extends from exospheric heights (from $10^4$ km altitude for the ionospheric flux tubes) down to 15 km altitude. The extensions to CTIP mean that lower atmosphere dynamic effects such as gravity waves can be included, and conversely the effects of ionospheric inputs such as auroral precipitation on middle and lower atmosphere can be examined. 

A Thermosphere General Circulation Model (TGCM) family, developed at the National Center for Atmospheric Research by Richmond et al. \cite{richmond92} comprises three-dimensional, time-dependent modules representing the Earth's neutral upper atmosphere. Recent models in the series include a self-consistent aeronomic scheme for the coupled Thermosphere/Ionosphere system, the Thermosphere Ionosphere Electrodynamic General Circulation Model (TIEGCM), and the TIME-GCM, which extends the lower boundary to 30 km and includes the effects of the prevailing physical and chemical processes.

Optical phenomena such as lightning-induced sprites, jets and elves and the 
electromagnetic fields associated with them have become a topic of intense study 
over the last decade. They are of too small a scale to be handled properly by 
global circulation and coupling models. This kind of electromagnetic activity 
is discussed in a companion chapter by E. Blanc.


\section{Can we predict future solar activity ?}
\label{geosci_future}

Solar activity is unfortunately very difficult to predict on time-scales that are most relevant for climate studies. The three time-scales that are most relevant for Sun-climate connections are associated with solar rotation (month), solar cycle (decade), several solar cycles (century). 

\subsection{Monthly variations}

Monthly variations are essentially associated with solar rotation (the synodic rotation period of the Sun is 27.5 days \footnote{Solar rotation is differential, so the period is somewhat longer at the poles than at the equator.}) and the emergence and waning of active regions. The first one is deterministic whereas the second one is more difficult to predict. Active regions can emerge in a few days, whereas their decay is generally more progressive. Predicting the emergence of active is today a major challenge, even though local helioseismology offers the possibility to anticipate subsurface changes in the magnetic field \cite{gizon05}.

Solar rotation in principle provides a convenient means for anticipating variations 14 days ahead, by assuming that the active regions do not evolve significantly in the meantime. This is often a reasonable assumption and indeed most solar activity indices exhibit a conspicuous 27-day periodicity, see Fig.~\ref{fig_indices}. The problem, however, is present lack of direct observations of the back side of the Sun. The twin Stereo spacecraft will soon be the first to witness the far side but no permanent monitoring is foreseen. There remains nevertheless a possibility to detect the presence active regions on the far side indirectly either by using time-distance helioseismology \cite{zhao07} or by using the interplanetary glow in the Lyman-$\alpha$ line \cite{quemerais02}. These techniques do not provide radiometrically accurate data, yet they are helpful for determining trends.

\subsection{Decadal variations}

The decadal time-scale is principally associated with the sunspot cycle, whose period varies from 10 to 13 years. This solar cycle actually represents half of a  rectified solar magnetic (or Schwabe) cycle. Numerous techniques have been developed for predicting the amplitude and the duration of a solar cycle, see for example the review by Hathaway \textit{et al.} \cite{hathaway99}. Once a cycle has started, the maximum amplitude can be predicted with an accuracy of about 10 \%, using empirical or semi-empirical criteria. Predicting a cycle before its onset, however, is a much more challenging task that requires a good understanding of the flows in the solar interior. Major progress has been made by running flux transport dynamo models   driven with sunspot data and calibrated with helioseismic inputs \cite{dikpati06}. Self-consistent dynamo models, however, are still out of reach.  

The lack of predictability of the solar cycle is well illustrated by the present lack of consensus on what the amplitude of the 24th solar cycle (which is expected to start in mid-2009) should be \cite{pesnell08}. Another reason stems from the chaotic nature of the solar dynamo (e.g. \cite{letellier06}), which by definition precludes long-term forecasts. It is therefore very unlikely that the 22-year prediction horizon associated with the magnetic cycle will be significantly extended in a near future.

\subsection{Centennial variations}

Unfortunately, the most relevant scale for Sun-climate studies is also is the least predictable one. No model can today predict solar activity several cycles ahead. From a statistical point of view, however, we know that the Sun spends about 15 \% of its time in periods of reduced activity, known as grand minima, during which the dynamo can stay for several decades or up to two centuries in a special state \cite{usoskin08}. The Sun also has states of high activity, called grand maxima. The second half of the 20th century arguably was a grand maximum \cite{solanki04}, whereas the Maunder and Sp\"orer periods were grand minima. Usoskin \cite{usoskin08} identified  at least 27 grand minima since 10000 BP. These results are all based on the analysis of cosmogenic isotopes such as $^{10}$Be and $^{14}$C \cite{bard06}. 

Comparisons with Sun-like stars \cite{baliunas90, lockwood92} show that grand minima are by no means restricted to the Sun only. Interestingly, these comparisons with solar analogues provide a useful means for setting lower and upper bounds to what the TSI may have been like in the past. These minima pose a major challenge since we do not know in what state the Sun exactly was. The density of the solar wind most likely was low, auroras were infrequent \cite{silverman92} and only few sunspots occurred erratically. Terrestrial records, however, still show the signature of a 11-year cycle during the Maunder and Sp\"orer minima \cite{miyahara07}, which suggests that the conspicuous variability associated with the presence of sunspots may not be the sole solar cause of climate change. A better understanding of these grand minima is probably the key to the quantification of long-term solar variability.

Reconstructions of past solar activity show that the onset of grand minima or maxima is unpredictable. These phases are known to occur rapidly, within a few years only. Their recovery, however, is more gradual and takes one or two solar cycles. 

\subsection{Orbital time scales and beyond}

Longer term predictions become quite predictable again, because one of the dominant mechanisms affecting solar variability is the orbital variation of the Earth and the Sun \cite{berger88, crucifix06}. These mechanisms are further discussed in the chapter by D. Paillard. 

On time scales from $10^6$ to $10^9$ years, the dominant mechanisms are the motion of our solar system through the galactic spiral arms \cite{shaviv02} (thus affecting the cosmic ray flux) but also the thermonuclear burning cycle in the solar interior. While the former is still a matter of debate, the second presently causes a slow increase in the irradiance, which is well described by the solar standard model \cite{stix02}, constrained by helioseismic data.


\section{Conclusions}
\label{geosci_conclusions}

Solar radiation is by far the most intense source of energy supplied to the terrestrial atmosphere, and there is a wealth of evidence in favour of the response of atmospheric parameters to solar variations. Most of the attention has focused so far on the sole variability of the total solar irradiance, which gives a simplistic view of the complexity of the solar driver. Indeed, solar variability manifests itself in a variety of different (but coupled) mechanisms; most of the underlying feedback mechanisms remain poorly known, which hampers the quantification of individual processes. For that reason, there has been and is still much debate about the real impact of solar variability on climate. According to the IPCC \cite{ipcc07b}, over the last century this impact has  most likely been small as compared to anthropogenic effects. 

There are several important working fronts as far as the Sun--Earth connection is concerned.  Most GCMs whose development started in the lower atmosphere still largely ignore the upper part of the atmosphere on which solar variability has the largest impact. One obvious issue is therefore the upward extension of these models, and a better description of the mechanisms by which the upper layers may couple to the stratosphere and eventually to the troposphere. This also involves a better understanding on how solar variability affects regional climate data. On the other hand, GCM models like the CITP which started from the thermosphere, face the challenge of an appropriate downward extension to the stratosphere (and eventually the troposphere).

A second issue  is the definition of reference spectral irradiance in the EUV and UV bands for different levels of solar activity. These bands have an important leverage of the middle atmosphere and the reconstruction of past levels is still lacking today. In all these reconstruction attempts, however, one should be careful against inbreeding of models.

A third issue is the understanding of the microphysics associated with atmospheric electricity and in particular the quantitative role of ions and electrons for stimulating the production of water vapour condensation nuclei. All three issues involve a much closer interaction between the space and atmospheric communities, which is definitely the highest priority of all.

\bigskip

\textbf{Acknowledgement.}
During the preparation of this paper JW benefited from a grant provided by 
Le Studium, \textit{agence r\'{e}gionale de recherche et d'accueil international de chercheurs associ\'{e}s en r\'{e}gion Centre}, France. We also gratefully acknowledge the numerous institutes and teams that provided the experimental data we used in the plots.


\bibliographystyle{abbrv}


\end{document}